\documentclass[aps,prl,twocolumn,showpacs,amsmath,amssymb,superscriptaddress]{revtex4-1}
\usepackage{graphicx}
\usepackage{dcolumn}
\usepackage{epstopdf}
\usepackage{bm}
\usepackage{color}

\begin{document}
\title{Quasiparticle band structure and tight-binding model for single- and bilayer black phosphorus}

\author{A.~N. Rudenko}
\email[]{a.rudenko@science.ru.nl}
\author{M.~I. Katsnelson}
\affiliation{Institute for Molecules and Materials, Radboud University Nijmegen, Heijendaalseweg 135, 6525 AJ Nijmegen, The Netherlands}
\date{\today}

\begin{abstract}
By performing \emph{ab initio} calculations for one- to four-layer
black phosphorus within the $GW$ approximation, we obtain a significant difference in the band gap ($\sim$1.5 eV),
which is in line with recent experimental data.
The results are analyzed in terms of the constructed four-band tight-binding model, which gives accurate
descriptions of the mono- and bilayer band structure near the band gap, and reveal an important 
role of the interlayer hoppings, which are largely responsible for the obtained gap difference.
\end{abstract}

\pacs{73.22.-f, 74.20.Pq, 71.10.Fd}

\maketitle

\emph{Introduction.}
Black phosphorus (BP) is a layered material consisting of puckered atomic layers of elemental phosphorus coupled together 
by weak van der Waals forces \cite{Morita}. BP is attracting attention because of the prediction of phosphorus nanotubes \cite{Seifert,Cabria} and especially in view of recent 
success in obtaining a few-layer BP,
broadening the range of two-dimensional (2D) materials \cite{Li,Liu,Koenig,Buscema,Gomez}. Preliminary investigations indicate a strong contrast in the
electronic properties of bulk and few-layer BP, giving rise to the possibility of novel practical applications \cite{Li,Liu,Koenig,Buscema,Gomez}.

Since high-quality BP crystal became available \cite{Lange}, the electronic properties of BP have been extensively studied experimentally.
In particular, bulk BP has been shown to be a semiconductor with a moderate band gap of 0.31--0.35 eV \cite{Keyes,Warschauer,Maruyama},
whereas liquid He temperatures
along with high pressure give rise to superconductivity \cite{Kawamura}. Despite containing only one $p$ element, a theoretical 
description of BP turns out to be very challenging. Earlier attempts could not provide a reliable description of the band structure
due to shortcomings of the computational methods \cite{Takao,Goodman,Asahina,Nolang}. Although the employment of more accurate
nonempirical approaches reported in recent studies yields more consistent results \cite{Du,Prytz}, their performance is strongly dependent on 
the quality of the exchange-correlation approximation.
 
In contrast to semiconducting bulk BP, monolayer BP is predicted to be an insulator with a considerably larger band gap, strongly
depending on the number of layers \cite{Takao,Asahina,Liu,Du,Rodin,Qiao}, which is also supported by experimental observations \cite{Buscema,Gomez}.
However, the origin of a considerable band gap broadening in going from bulk to monolayer remains unclear.

In this Rapid Communication, we analyze in detail the electronic properties of monolayer, multilayer ($n$=2--4), and bulk BP within the quasiparticle $GW$ approximation. 
Particularly, we address the problem of the variation of their electronic properties. To this end, we construct a tight-binding model, which sheds light on the mechanism of the band gap 
formation in BP and further can be used in large-scale calculations of transport and optical properties.

\begin{figure}[t]
\includegraphics[width=0.47\textwidth, angle=0]{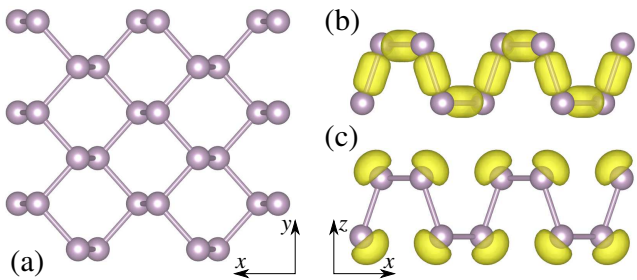}
\caption{(Color online) (a) Top view of the crystal structure of monolayer BP, and
side views of the occupied orbitals, corresponding to (b) bonding orbitals
and (c) lone pairs. The orbitals are given in terms of the 
maximally localized Wannier functions \cite{wannier90,Marzari} obtained in this work.}
\label{structure}
\end{figure}

\emph{Structure and chemical bonding.}
A single layer of BP consists of a corrugated arrangement of P atoms and has a thickness of $\sim$5~\AA~[Fig.~\ref{structure}(a)] \cite{Brown,Cartz}.
Alternate stacking of the layers along the [001] direction gives rise to the structure of bulk BP, which is stabilized by weak dispersive 
interactions. 
The intralayer bonding in BP results from the 
$sp^3$ hybridization of P atoms, giving rise to three bonding orbitals per two atoms [Fig.~\ref{structure}(b)] augmented by lone pairs associated 
with each atom [Fig.~\ref{structure}(c)]. The latter plays a particular role in the pressure-induced transformations of BP, as well as accounts
for a variety of structural modifications of solid P \cite{Boulfelfel}.

\emph{Electronic structure and a band gap in bulk BP.}
We first calculate the band structure of bulk BP along the high-symmetry lines
of the Brillouin zone (BZ) by using two different theoretical approaches.
The first method is the standard generalized-gradient 
approximation (GGA) \cite{pbe} that is routinely used in density functional theory (DFT) calculations, 
while the second one corresponds to an explicit calculation of the self-energy ($\Sigma=iGW$) within 
the $G_0W_0$ procedure \cite{Hedin,Shishkin}, where both the Green's function $G_0$ and screened 
exchange $W_0$ are evaluated using DFT-GGA wave functions. 

The calculations presented in this work were
carried out by using the Vienna \emph{ab initio} simulation package ({\sc vasp}) \cite{kresse1996,vasp_paw}.
An energy cutoff of 280 eV for the plane-wave basis and the convergence threshold of 10$^{-8}$ eV were employed to 
obtain the DFT wave functions. The number of unoccupied bands in $GW$ calculations was set
to 90 per atom and 70 grid points were used for integration along the frequency axis. To sample the Brillouin zone, 
{\bf k}-point meshes of (10$\times$12$\times$4) and (10$\times$12$\times$1) were used for bulk and multilayer
calculations, respectively. An experimental lattice structure was adopted in all cases \cite{Brown,Cartz}. For slab
(multilayer) calculations, a vacuum layer of $\sim$20~\AA~was used. The chosen set of parameters ensures
that the one-particle energies are accurate to within a few tens of meV.

\begin{figure}[b]
\includegraphics[width=0.49\textwidth, angle=0]{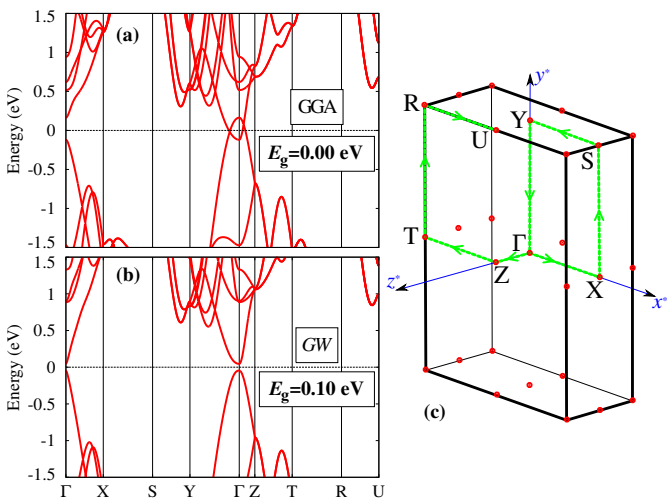}
\caption{(Color online) Band structures of bulk BP calculated by using (a) the DFT-GGA and (b) a more accurate 
$GW$ approach along the high-symmetry points of the Brillouin zone. The corresponding path is shown by a green 
line in (c). Zero energies in (a) and (b) correspond to the Fermi level (GGA) and center of the band gap ($GW$).}
\label{bulk_bands}
\end{figure}

In Fig.~\ref{bulk_bands}, we show the band structure of bulk BP calculated by using the two different methods. 
One can see that both GGA and $GW$ band structures exhibit similar features with the exception of the
relative position of the valence (VB) and conduction (CB) bands, which results in different band gaps ($E_\mathrm{g}$). In particular, the GGA approach 
leads to an overlap between the VB and CB in the vicinity of the $\Gamma$ point (zero band gap), whereas the $GW$ method gives rise
to a band gap of $\sim$0.1 eV. Although both approaches do not reproduce the experimental band gap of 0.31--0.35 eV, the $GW$ method
yields a qualitatively correct trend toward the band gap opening, which is expected to be improved by a self-consistent treatment
of $G$ and $W$. Being in principle possible, such a treatment is highly demanding computationally and not considered within the 
present work. A qualitative difference between the results of the DFT-GGA and $GW$ methods indicates an important role of electron
correlations in BP, which requires a careful theoretical treatment.

To analyze the orbital composition of the bands close to the gap, we project the $GW$ quasiparticle states onto the canonical $s$ and 
$p_i$ ($i=x,y,z$) orbitals, which allows us to decompose the VB and CB into different orbital contributions.
The decomposition at the $\Gamma$ point yields
$\left| \psi^{\mathrm{VB}} (\Gamma) \right>=0.17\left| s \right> + 0.40\left| p_x \right> + 0.90\left| p_z \right>$
and 
$\left| \psi^{\mathrm{CB}} (\Gamma) \right>=0.57\left| s \right> + 0.44\left| p_x \right> + 0.69\left| p_z \right>$,
respectively, for VB and CB. One can see that the relevant bands represent a mixture of all the orbitals, with the exception of $p_y$
having zero contribution at $\Gamma$. Although the $p_z$ orbital has the largest contribution in both cases, the role of the other orbitals
($s$ and $p_x$) in the formation of VB and CB cannot be considered as negligible.
Therefore, we emphasize that in contrast to graphite (graphene), whose relevant bands are determined exclusively by the $p_z$ states, 
the band structure of BP is considerably less trivial due to the mixture of states of different symmetry.

\begin{figure*}[t]
\includegraphics[width=0.75\textwidth, angle=0]{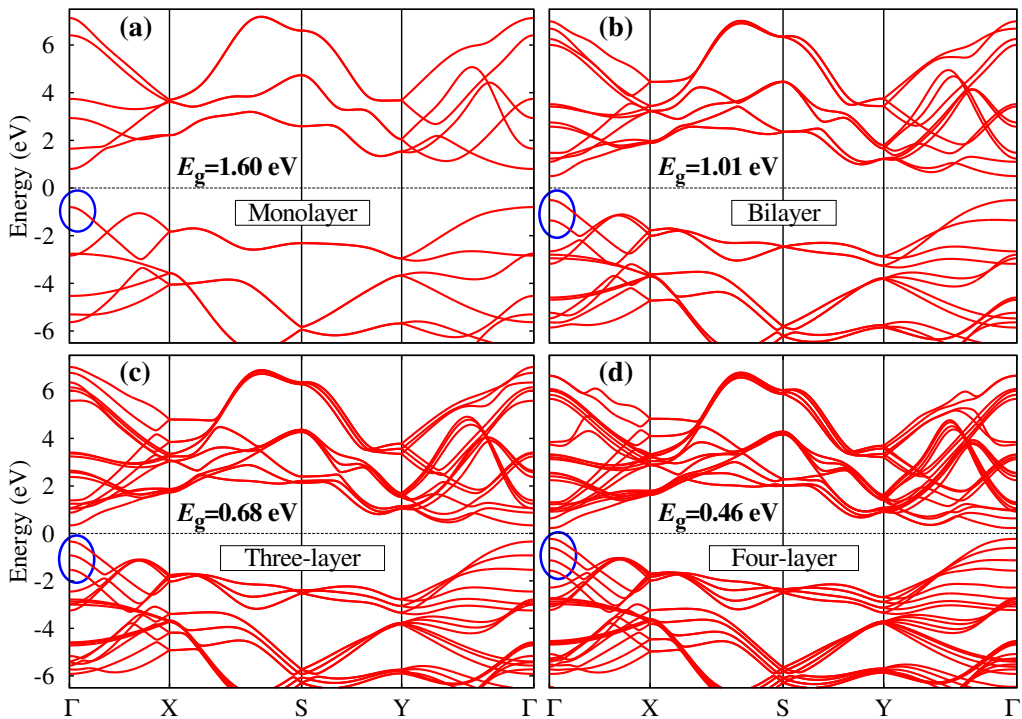}
\caption{(Color online) Band structures for $n$-layer BP calculated within the $GW$ approach for $n$=1--4. Zero energy corresponds to the center of the band gap.
Blue circles show the band splitting near the gap.}
\label{multilayer_bands}
\end{figure*}

\emph{Band structures of monolayer and a few-layer BP.}
As a next step, we apply the $GW$ approximation to the calculation of the band structure of monolayer and a few-layer BP.
In Fig.~\ref{multilayer_bands}, we show the corresponding spectra for a different number of layers ($n$=1--4). One can see that
in the case of the monolayer, all spectral features remain essentially the same as for bulk BP, with the exception of the gap between the VB and CB,
which also appears at the $\Gamma$ point, but has a significantly higher value (1.60 eV). The addition of more layers results in
the band splitting over the entire BZ, which, in turn, leads to the decrease of the gap. The band gap decreases monotonically
with the number of layers, reaching the value of 0.46 eV in the four-layer case.
The observed trend is in line with previous DFT investigations \cite{Du,Qiao,Liu},
although the $GW$ approach results in an appreciably larger band gap. It should be noted that although recent hybrid-functional DFT calculations 
within the HSE06 scheme \cite{hse06} report a similar band gap for the monolayer BP (1.51 eV) \cite{Qiao}, the application of the same approach to
bulk BP leads to a substantial overestimation of its band gap (0.82 eV) \cite{Liu}, whereas the adjustment of the functional to give a better description for
bulk BP conversely reduces the monolayer values ($\sim$1.16 eV).

Existing experiments on the photoresponse of BP-based field-effect transistors provide an estimation of the cutoff wavelength for the excitation of 
the carriers of a few-layer BP, which amounts to 1.24 eV \cite{Buscema}.
Similarly, photoluminescence measurements provide indications of an even larger optical gap of 1.6 eV \cite{Gomez}.
We note that a direct comparison between theory and experiment is not possible since the number of layers in the experimental samples is not clearly determined, 
while the $GW$ approach does not capture excitonic effects, which are necessary for a correct description of the optical spectra. Nevertheless,
the experimentally reported values can be considered as a lower limit for the band gap in monolayer BP, which indeed indicates that the gap in BP
is strongly dependent on the number of layers. In comparison with previous DFT studies, the $GW$ results presented above are appreciably closer to
experimental observations. Moreover, taking into account some underestimation of the $GW$ band gap in bulk BP, the same trend is expected for monolayer 
and multilayer BP, which suggest that the actual monolayer band gap is larger than the obtained value of 1.60 eV.

\emph{Tight-binding parametrization.}
We now turn to the tight-binding (TB) analysis of the band structure.
Previously, a two-band model has been proposed within the {\bf k}$\cdot${\bf p} approximation for monolayer
BP \cite{Rodin}, which yields a reasonable description of the DFT bands near the $\Gamma$ point. 
However, the effective Hamiltonian proposed in Ref.~\onlinecite{Rodin} is determined in reciprocal space and
does not involve any real-space interaction parameters, which is necessary to have an insight into the origin 
of the gap and its evolution with the number of layers.

Here, we further analyze the electronic structure of monolayer BP by performing TB parametrization of the $GW$ 
Hamiltonian by using the following four-band model,
\begin{equation}
H=\sum_{i} \varepsilon_i n_i + \sum_{i\neq j} t^{||}_{ij} c_i^{\dag}c_j,
\label{tb_hamilt}
\end{equation}
where the summation runs over the lattice sites of single-layer BP (four sites per unit cell), $\varepsilon_i$ is the energy of the electron at site $i$, $t^{||}_{ij}$ is the 
hopping parameter between the $i$th and $j$th sites, and $c^{\dag}_i$ ($c_j$) is the creation (annihilation) operator of electrons at site $i$ ($j$).
To obtain an effective Hamiltonian in the form of Eq.~(\ref{tb_hamilt}), we first construct a set of four maximally localized Wannier functions $\left|w_i({\bf r})\right>$ \cite{wannier90,Marzari} 
by freezing
the states in the region of 0.3 eV above and below the band gap. We then obtain the matrix elements of the original $GW$ Hamiltonian in the Wannier function basis $\left<w_i|H|w_j\right>$, which
can be directly associated with the $\varepsilon_i$ and $t^{||}_{ij}$ parameters appearing in Eq.~(\ref{tb_hamilt}). Finally, we cut less significant parameters by employing the criteria $|t^{||}_{ij}|<0.1$ eV, 
and reoptimize the remaining parameters in order to obtain a better band description within the relevant energy region.

    \begin{table}[!bt]
    \centering
    \caption[Bset]{Inlayer ($t^{||}$) and interlayer ($t^{\perp}$) hopping parameters obtained in terms of the TB Hamiltonian [Eqs.~(\ref{tb_hamilt}) and (\ref{tb_bilayer})] for monolayer and bilayer BP. 
$d$ and $N$ denote the distances between the corresponding interacting lattice sites and the coordination number at the given distance, respectively. The hoppings are schematically shown in 
Fig.~\ref{model_bands}(c).}
    \label{hoppings_table}
 \begin{tabular}{ccccccccc}
      \hline
      \hline
 &  \multicolumn{3}{c}{Inlayer} & & \multicolumn{3}{c}{Interlayer} \\
\cline{2-4}
\cline{6-8}
      No.      & \, $t^{||}$, eV \, & \, $d_{||}$, \AA \, & \, $N^{||}$  \, &      & \, $t^{\perp}$, eV \, & \, $d_{\perp}$, \AA \, & \, $N^{\perp}$ \, \\
     \hline
       1       &  \   $-$1.220  \   &  \    2.22  \   &  \   2   \      &      &   \  \,\,\,\,\,0.295  \     & \      3.60   \  & \  2 \ \\
       2       &  \  \,\,\,\,\,3.665  \   &  \    2.24  \   &  \   1   \      &      &   \  \,\,\,\,\,0.273  \     & \      3.81   \  & \  2 \ \\
       3       &  \   $-$0.205  \   &  \    3.34  \   &  \   2   \      &      &   \ $-$0.151  \     & \      5.05   \  & \  4 \    \\
       4       &  \   $-$0.105  \   &  \    3.47  \   &  \   4   \      &      &   \ $-$0.091  \     & \      5.08   \  & \  2 \    \\
       5       &  \   $-$0.055  \   &  \    4.23  \   &  \   1   \      &      &   \  \,\,\,\,\,0.000  \     & \      5.44   \  & \  1 \ \\
      \hline
      \hline
    \end{tabular}
    \end{table}

\begin{figure*}[t]
\includegraphics[width=0.75\textwidth, angle=0]{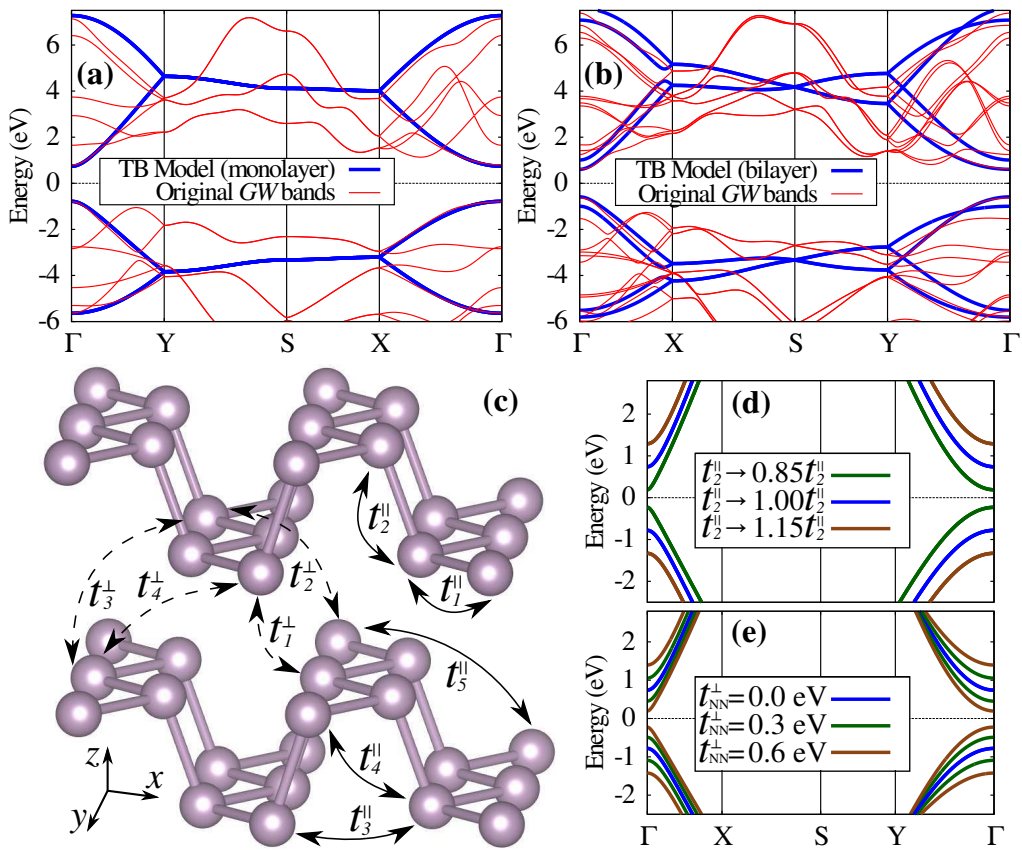}
\caption{(Color online) Band structures calculated by using the tight-binding parametrization (see text for details) in comparison with the original $GW$ bands for (a) monolayer and (b) bilayer BP. Hopping parameters of the TB model are sketched in (c). (d) and (e) show the dependence of the monolayer TB model on the inlayer ($t^{||}_2$) and nearest-neighbor interlayer ($t^{\perp}_{\mathrm{NN}}$) hopping parameters.}
\label{model_bands}
\end{figure*}

In Table I, we list the obtained TB parameters for monolayer BP, which is described by five inlayer hoppings up to a
distance of 4.23~\AA~[see Fig.~\ref{model_bands}(c)]. We note that due to symmetry, the electron energies ($\varepsilon_i$) 
appearing in Eq.~(\ref{tb_hamilt}) are equivalent for all lattice sites.
The corresponding model band structure 
is shown in Fig.~\ref{model_bands}(a) in comparison with the original $GW$ bands. One can see that both electron and hole states are accurately reproduced within the region of $\sim$0.3 eV each. Beyond 
that region, the four-band model does not give a reliable description due to the presence of additional bands of different symmetry. As can be seen from Table \ref{hoppings_table}, the
band structure of monolayer BP is determined predominantly by the first two parameters, which describe the nearest-neighbor in-plane ($t^{||}_1$) and nearest-neighbor 
out-of-plane ($t^{||}_2$) hoppings in the system. Apart from being positive, $t^{||}_2$ has the largest magnitude, which indicates a particularly important role of this parameter in 
determining the electronic structure. In Fig.~\ref{model_bands}(d), we show the TB bands in the vicinity of the band gap calculated by varying the $t^{||}_2$ parameter. One can notice that
the increase (decrease) of $t^{||}_2$ results in a uniform shift of the VB and CB toward (apart from) each other, while the shape of the bands remain unchanged. Not only can such a behavior
be used for the adjustment of the band gap in monolayer BP, but it also points to a significant role of the intersite Coulomb repulsion ($V$) between $p$ electrons in BP. This observation
is consistent with the improper treatment of the Coulomb repulsion within the DFT, leading to the well-known underestimation of band gaps in insulators and semiconductors. In addition, we note
that there is no direct hopping between the nearest-neighbor sites along the $y$ direction, which accounts for the highly anisotropic transport properties in BP \cite{Xia,Fei}.

Let us now consider the case of two single-layer BP with additional interlayer hoppings $t^{\perp}$. The TB Hamiltonian reads
\begin{equation}
H=\sum_{i} \varepsilon_i n_i + \sum_{i\neq j} t^{||}_{ij} c_i^{\dag}c_j + \sum_{i \neq j} t^{\perp}_{ij}c_i^{\dag}c_j,
\label{tb_bilayer}
\end{equation}
where the first two terms have the same meaning as in Eq.~(\ref{tb_hamilt}), whereas the third term describes interactions between sites belonging to different layers.
We first analyze the situation where only the nearest-neighbor interlayer hoppings ($t^{\perp}_{\mathrm{NN}}$) are present.
In Fig.~\ref{model_bands}(e), we show the relevant part of the band structures calculated for different values of $t^{\perp}_{\mathrm{NN}}$. 
One can see that the interlayer hopping leads to the splitting of the bands near the $\Gamma$ point, whose broadening is
proportional to $t^{\perp}_{\mathrm{NN}}$. Therefore, the reduction of the band gap in multilayer BP can be qualitatively described just by introducing the interlayer hopping parameter.

The relevant part of the bilayer band structure can be more consistently reproduced by using the same set of inlayer hopping parameters augmented by an energy splitting $\Delta \varepsilon=1.0$ eV
between the energies of nonequivalent electrons and four interlayer hoppings, listed in Table \ref{hoppings_table}. The corresponding model bands for the bilayer are shown in 
Fig.~\ref{model_bands}(b). As in the case of the monolayer, the bilayer model yields an accurate description of the bands in the vicinity of the gap. However, the band splittings clearly visible
in the original band structure (see Fig.~\ref{multilayer_bands}) are largely underestimated. On the other hand, a better reproduction of the band splitting would apparently lead to a significantly 
reduced band gap, worsening
the agreement with the original results. Such behavior indicates that a more reliable TB model for bilayer and multilayer BP is supposed to also take into account the change of the inlayer hoppings 
parameters $t^{||}$.

Recently, we became aware of the recent work of
Tran \emph{et al.} \cite{Tran} The
authors of Ref.~\onlinecite{Tran} also report the results of a non-self-consistent 
$GW$ calculation for a few-layer BP ($n$=1--3), but with significantly
higher energy gap values obtained ($\sim$2.0 eV in the monolayer case). An apparent 
inconsistency with the results of the present work may be explained by the
use of the general plasmon pole model for approximating the screened Coulomb
interactions $W$ in Ref.~\onlinecite{Tran}, in comparison with a complete
random phase approximation (RPA) \cite{Shishkin} in our work.
The difference in the results emphasizes once more an important role 
of the screening effects in BP.

\emph{Conclusions.}
By performing quasiparticle $GW$ calculations, we have shown that the band gap in black phosphorus is strongly dependent on the number of layers, yielding 1.6 and 0.1 eV for the monolayer and
bulk cases, respectively. The origin of the band gap has been analyzed in terms of a four-band tight-binding model.
In contrast to graphene, where one nearest-neighbor hopping parameter only is sufficient to reproduce the main characteristics of the energy spectrum \cite{Katsnelson-Book}, the minimal model for a 
single-layer BP involves two important parameters, describing the in-plane ($t^{||}_1$=$-$1.22 eV) and out-of-plane ($t^{||}_2$=3.67 eV) nearest-neighbor hoppings. Moreover, the
appearance of the second (repulsive) parameter, which is a consequence of the puckered BP structure, is shown to be largely responsible for the band gap opening. The reduction of the band gap with
the number of layers can be qualitatively described by introducing the repulsive hopping parameters ($t^{\perp}$) between the layers. An accurate description, however, does not appear
possible without taking the change of the inlayer hoppings ($t^{||}$) into account, which is not typical for other known two-dimensional materials, particularly graphene \cite{Katsnelson-Book} and 
metal dichalcogenides \cite{Cappelluti}.

\emph{Note added in proof.}
Recently, a number of model calculations on
the optical and transport properties of monolayer and a few-layer BP have been reported \cite{Low1,Low2,Ezawa}. In particular, 
the recent work of Ezawa \cite{Ezawa} is based on the model proposed in the present Rapid Communication, providing an example 
of its application.

\emph{Acknowledgement.}
The authors are thankful to Kostya Novoselov for stimulating discussions.
The research has received funding from the European Union Seventh 
Framework Programme under Grant Agreement No.~604391 Graphene Flagship.

\end{document}